\documentclass[12pt]{article} 

\usepackage{color}
\usepackage{xcolor}
\usepackage{pdfsync}
\usepackage{amssymb}
\usepackage{amsmath,amsthm,amsfonts}  
\usepackage{graphicx}
\usepackage[breaklinks,colorlinks,backref]{hyperref}

\begin{document}

\title{
{\Large \bf A Nonlinear $q$-Deformed Schr\"odinger Equation}
}

\author{
M. A. Rego-Monteiro$^1$ and E. M. F. Curado$^{1,2}$  \\
$^1$Centro Brasileiro de Pesquisas F\'\i sicas \\ 
$^2$ Instituto Nacional de Ci\^encia e Tecnologia de Sistemas Complexos \\
 Rua Xavier Sigaud 150, 22290-180 \\ Rio de Janeiro, RJ, Brazil \\
e-mails:   regomonteirojgd@gmail.com, evaldo@cbpf.br }  
\maketitle

%

\begin{abstract}
 
\indent

We construct a new nonlinear deformed Schr\"odinger structure using a nonlinear derivative operator which depends on a parameter $q$. This operator recovers Newton derivative when $q \rightarrow 1$. Using this operator we propose a deformed Lagrangian which gives us a deformed nonlinear Schr\"odinger equation with a nonlinear kinetic energy term and a standard potential $V(\vec{x})$. We analytically solve the nonlinear deformed Schr\"odinger equation for $V(\vec{x}) = 0$ and $q \simeq1$. This model has a continuity equation, the energy is conserved, as well as the momentum and also interacts with electromagnetic field. Planck relation remains valid and in all steps we easily recover the undeformed quantities when the deformation parameter goes to 1. Finally, we numerically solve the equation of motion for the free particle in any spatial dimension, which shows a solitonic pattern when the space is equal to one for particular values of $q$.


\end{abstract}

\begin{tabbing}

\=xxxxxxxxxxxxxxxxxx\= \kill

 Nonextensive thermostatistics; 
Solitons;  $q$-derivative.

\\

\end{tabbing}

\newpage

\section{Introduction}
\label{intro}

 The Nonlinear Schr\"odinger Equation (NLSE) is an important model that arises from introducing a nonlinear term into the standard Schr\"odinger equation. First developed within 1+1 dimensional field theory \cite{zakharov}  and later generalized 
\cite{iwo}, the NLSE provides a powerful framework for understanding phenomena in diverse fields including nonlinear optics, plasma physics \cite{agraval}, Bose-Einstein condensation \cite{pitaevskii},  water waves \cite{sulem} and continuing to arouse much interest to this day \cite{recentes}. In this work, we present a fundamentally different approach to nonlinearizing the Schr\"odinger equation. Instead of modifying the potential, we introduce a nonlinearity in the kinetic energy term. We identify this new model as the qNLSE.

There is yet another purpose of constructing a nonlinear variation of quantum mechanics. The idea that the standard quantum mechanics we know today can  be an approximation of a more general nonlinear theory has its origins a long time ago. According to standard textbooks de Broglie and Einstein believed that quantum mechanics was incomplete and was an approximation of a more general theory \cite{croca}. Attempts to build nonlinear quantum mechanics date back a long time ago (see for instance \cite{croca} and references there in). For instance, Guerra and Pusterla \cite{guerrapusterla}, Smolin \cite{smolin}, also Gueret and Vigier \cite{gueretvigier}, and others, developed in the eighties nonlinear approaches to quantum physics. However, we must say that no one has come close to giving a satisfactory answer to all the difficulties of this purpose. To this day we do not have a satisfactory proposal for a nonlinear theory of quantum mechanics, and neither is it the objective of this work. The purpose of this work is to perform a much more simple program.

 What in the literature we call $q$-deformation refers to the introduction of one or more parameters into a physical model or mathematical structure in such a way that when these parameters assume a certain value, we recover a known physical model or mathematical structure (\cite{kulish}, \cite{sklyanin}, \cite{drinfeld}, \cite{jimbo}). This generalization was also constructed with derivative operators that recovered Newton's derivative, giving rise to a field in mathematics known as q-calculus \cite{q-calculus}. Later, for instance, one of these possible operators, the Jackson derivative operator, was used to construct a deformed linear quantum mechanics \cite{lavagno}.

In \cite{nrt}, a deformed version of the Schr\"odinger equation for a free particle where a nonlinearity is present in the kinetic term of this equation was proposed. 
The interest on this generalized nonlinear wave equation was motivated, in part, by its connection with a class of nonlinear Fokker-Planck equations, which is related to the 
nonextensive statistical mechanics \cite{tsallis1,tsallisbook, CuradoNobre2003}. 
A nonlinear (power-law) Fokker-Planck equation has as solution the 
so-called $q$-exponential, which is a distribution presenting a power-law tail for larger values of the 
variable (like the energy).  Actually, it can be proved by means 
 of a $H$-theorem \cite{VeitNobreCurado2007, VeitCuradoNobre2007} that the $q$-exponential 
 solution of the nonlinear Fokker-Planck equation is a stable solution. 
 But, this same distribution can be obtained by the extremization of a particular 
entropic form $S_q$, which depends on the same parameter $q$ \cite{tsallisbook}. This shows 
 the deep relation exhisting between the nonlinear Fokker-Planck equation, 
 the $S_q$ entropy and the framework of the nonextensive statistical mechanics.  
Several works have shown the consistency of this generalized statistical mechanics and its usefulness to describe some physical systems (see, for instance \cite{tsallisbook}). 
In fact, as was shown in \cite{nrt}, the 
generalized nonlinear Schr\"odinger equation for a free particle also has as solution the $q$-exponential distribution as the nonlinear Fokker-Planck equation.

In an optical fiber, NLSE describes the propagation of the light pulse envelope. The nonlinearity originates from the response of the medium, mainly due to nonlinear behavior with the electromagnetic field, as 
shown by the dielectric constant.
It is interesting to study  
different types of nonlinearity that could be linked to different nonlinear responses of the medium. The approach chosen in this paper is to use a nonlinear derivative that was introduced in non-extensive statistical mechanics which has already shown to describe many physical systems \cite{tsallisoverview}.

Another motivation for this nonlinearization of the Schr\"odinger equation is to 
find models whose solutions are integrable in the whole space, i.e.,  realistic solutions, 
belonging to a HIlbert space, not diverging as is the plane wave solution.

In \cite{nrt2} it was shown that in order to describe correctly the classical field structure of the model described in \cite{nrt} an additional field $\Phi(\vec{x},t)$ must be introduced besides to the usual field $\Psi(\vec{x},t)$ . This additional field becomes $\Psi^{*}(\vec{x},t)$ when $q \rightarrow 1$. In addition, this model has no global gauge invariance and cannot interact with light for $q \neq 1$.

In this paper, we construct a nonlinear Schr\"odinger deformed structure where a different nonlinearity is present in the kinetic term using only one field $\Psi(\vec{x},t)$ and its complex conjugate. This model has the additional advantage that can interact with light. To achieve this purpose we have used a nonlinear derivative operator which is generalization of the nonlinear operator given in \cite{nrt}, which is described in section (\ref{NEDOP}). In subsection (\ref{lagrangianoequacao}) using this nonlinear derivative we propose a deformed Lagrangian and right away we present the qNLSE for this deformed Lagrangian. In subsection (\ref{gauge}) we discuss gauge invariance for this deformed Lagrangian, introduce covariant derivatives and present the Lagrangian which is invariant under local gauge invariance. In subsection (\ref{eqcontinuity}) we show that this model has an equation of continuity for every solution of the qNLSE.  To end this section, we present in subsection (\ref{perturbative}) the perturbative solution of the qNLSE to first order in $\epsilon = 1-q$.  In section (\ref{sechamiltonian}) we generalize the connection between the generalized coordinate associated to the discretized field $\Psi_i(t)$ and we compute the deformed Hamiltonian of the model. In subsections (\ref{conservationen}) and (\ref{conservationmom}) we prove the conservation of energy and momentum of the model, respectively. In section (\ref{numerical}) we present the results for the numerical computation of the equation of motion. Finally, in section (\ref{final}) we present the final comments.

\section{Non-Extensive Derivative Operator}
\label{NEDOP}

Much of the description we have of physical nature is through Newton's differential and integral calculus. But this is not the only way of describing physics. There is an area of mathematics 
known as $q$-calculus which addresses other approaches such as finite differences \cite{q-calculus}. Two of the most studied derivatives in this area are the Jackson derivative  and the $h$-derivative which are defined as
\begin{eqnarray}
\label{jackson}
D_q f(x) &=& \frac{f(qx)-f(x)}{(q-1)x} , \\
\label{hderivative}
D_h f(x) &=& \frac{f(x+h)-f(x)}{h} ,
\end{eqnarray}
respectively.
It is clear that
\begin{equation}
\label{limit}
\lim_{q \rightarrow 1} D_q f(x) = \lim_{h \rightarrow 0} D_h f(x) = \frac{d f(x)}{dx} ,
\end{equation}
if $f(x)$ is differentiable. It is well-known that the above operators are linear operators and that each different derivative operator has one or more than one, different generalizations of the exponential, called $q$ or $h$-exponential.

In 2011, it was introduced a nonlinear $q$-derivative operator (defined in the paragraph below eq. (13) in reference \cite{nrt})  as 
\begin{equation}
\label{generalizadodnrt}
\mathbb{D}^{(q)}_{x} f(x) \equiv \frac{1}{q} \frac{d f(x)^q}{dx} ,
\end{equation}
for a real parameter $q$. Notice that, $\mathbb{D}^{(q)}_{x}$ is a nonlinear power-law generalization of the Newton's derivative. Moreover, $\mathbb{D}^{(q)}_{x} f(x) \neq  D_q f(x) \neq D_h f(x)$. 
It easy to verify that the $a$-exponential associated to the operator in eq. (\ref{generalizadodnrt}) is
\begin {equation}
\label{aexponential}
e_q(x) = \left(  1+(q-1) x \right)^{\frac{1}{q-1}},
\end{equation}
since $\mathbb{D}^{(q)}_{x} e_q(x) = e_q(x)$. Notice that, when $q \rightarrow 1$ the derivative operator defined in (\ref{generalizadodnrt}) becomes the standard Newton's derivative as well as the $q$-exponential in (\ref{aexponential}) becomes the standard exponential $\exp(x)$. We are going to use in this paper the derivative $\mathbb{D}^{(q)}_{x}$ and also, in Section \ref{gauge}, and  the covariant derivatives $\vec{\mathcal{D}}$ and $\mathcal{D}_t$.



Finally, we generalize the standard gradient as
\begin{equation}
\label{qgradient}
\vec{\mathbb{\nabla}}^{(q)} f(x) \equiv (\mathbb{D}^{(q)}_{x}, \mathbb{D}^{(q)}_{y}, \mathbb{D}^{(q)}_{z}).
\end{equation}
Where the standard gradient, $\vec{\nabla}$, is obtained from $\vec{\nabla}^{(q)}_{x}$  as $\vec{\nabla}=\vec{\mathbb{\nabla}}^{(1)}$,  $\vec{\nabla}^{(q)}_{x}$ for $q=1$.

\section{Deformed Schr\"odinger Structure}
\label{q-Schrodinger}

There is a well-known NLSE which was introduced by V. E. Zakharov and S. V. Manakov  \cite{zakharov}
and after generalized to arbitrary power-law nonlinearities which we present the Lagrangian below
\begin{equation}
\label{manakov}
\mathcal{L} = i \hbar \Psi \partial_t \Psi  - \frac{\hbar^2}{2 m} \vec{\nabla} \Psi^{*} . \vec{\nabla}\Psi - \frac{G}{2} \left( \Psi^{*} \Psi  \right)^q - V(\vec{x}) \Psi^{*} \Psi ,
\end{equation}
where $q$ is an arbitrary real number. Notice that the power-law nonlinear term $\left( \Psi^{*} \Psi  \right)^q$ is added to the standard Schr\"odinger structure.

\subsection{Deformed Lagrangian and Schr\"odinger Equation}
\label{lagrangianoequacao}

In this paper we take another way in order to introduce the nonlinearity. Using the power-law nonlinear operator $\mathbb{D}^{(q)}$ discussed in the previous section and the standard $\vec{\nabla}$ operator we are going now to introduce the following Lagrangian density 
\begin{equation}
\label{qlagrangian}
\mathcal{L} = i \hbar \Psi^{* q} \mathbb{D}^{(q)}_t \Psi  - \frac{\hbar^2}{2 m} \vec{\nabla} \Psi^{*} . \vec{\nabla}\Psi - V(\vec{x}) \Psi^{*} \Psi,
\end{equation}
where $\Psi(\vec{x},t) = \psi(\vec{x},t)/\psi(0,0)$.
Notice that, since $\mathbb{D}^{(1)} $ is the standard derivative and gradient, by construction this Lagrangian reproduces the standard Schr\"odinger Lagrangian in the limit $q\rightarrow 1$. Of course, if one wishes one could explicitly rewrite the above Lagrangian in terms of the standard derivatives. Moreover, it is important to observe that using eq. (\ref{generalizadodnrt})  we can easily prove the reality of the above Lagrangian in the same way as it happens in the limit $q\rightarrow 1$.

Using the standard Euler-Lagrange equations for the field $\Psi^{*}$ we arrive at the following $q$-deformed Schr\"odinger equation
\begin{equation}
\label{qschrodinger2}
i\hbar \partial_t \Psi^q   +   \frac{\hbar^2 }{2  m} \Psi^{*(1-q)} \nabla^2 \Psi  - V(\vec{x}) \Psi^{*(1-q)} \Psi
= 0 .
\end{equation}

\subsection{Invariance Under Gauge Transformation}
\label{gauge}
In order to analyze the symmetries of the Lagrangian it is convenient rewrite eq. (\ref{qlagrangian}) as
\begin{equation}
\label{qlagrangian2}
\mathcal{L} = \frac{i \hbar}{q} \Psi^{* q} \partial_t \Psi^q  - \frac{\hbar^2}{2  m} \vec{\nabla} \Psi^{*} . \vec{\nabla} \Psi - V(\vec{x}) \Psi^{*} \Psi .
\end{equation}
Written in this way we trivially see that the above Lagrangian is invariant under the global transformation $\Psi' = e^{ie \theta} \Psi$, where $e$ is the electric charge. Now, we would like to construct a Lagrangian that is invariant under the local transformation
 \begin{equation}
\label{localgauge}
\Psi' = e^{ie \theta(\vec{x},t)} \Psi ,
\end{equation}
where now the parameter $\theta$ depends on time and space.

In order this to happen we introduce the covariant derivatives
\begin{eqnarray}
\label{coderivative1}
\vec{\mathcal{D}} &=& \vec{\nabla} -i e \vec{A}(\vec{x},t), \\
\label{coderivative2}
\mathcal{D}_t  &=& \partial_t + i e q A_t(\vec{x},t) ,
\end{eqnarray}
where $e$ is the electric charge. Notice that, differently from the standard case the coefficients in the spatial and temporal cases slightly differ depending on the parameter $q$. 

As $\Psi$ transforms as $\Psi' = e^{ie \theta(\vec{x},t)} \Psi$, the fields $\vec{A}(\vec{x},t)$ and $A_t(\vec{x},t)$ must transform as
\begin{eqnarray}
\label{fieldtransf1}
\vec{A}'(\vec{x},t) &=& \vec{A}(\vec{x},t) + \vec{\nabla} \theta(\vec{x},t), \\
\label{fieldtransf2}
A{_t}'(\vec{x},t) &=& A_t(\vec{x},t) - \partial_t \theta(\vec{x},t) .
\end{eqnarray}

In this way, the Lagrangian invariant under the local transformation in eq. (\ref{localgauge}) is
\begin{equation}
\label{lagrangianemfield}
\mathcal{L} = \frac{i \hbar}{q} \Psi^{* q} \mathcal{D}_t \Psi^q  - \frac{\hbar^2}{2 m} \vec{\mathcal{D}} \Psi^{*} . \vec{\mathcal{D}} \Psi- V(\vec{x}) \Psi^{*} \Psi . 
\end{equation}
 Using in the above Lagrangian the covariant derivatives given in eqs. (\ref{coderivative1}) and ({\ref{coderivative2}) we see that a charged particle interacting with the electromagnetic field is modified by the parameter $q$ as if the charge would be redefined as $q e$. Thus if the proposal of this model were to be used to describe a quantum particle the value of $q$ should be very close to one.

\subsection{Equation of Continuity}
\label{eqcontinuity}

The procedure for calculating the equation of continuity is the same as in the standard Schr\"odinger case. Firstly, we generalize the probability density as $\rho = (\Psi^{*} \Psi)^q$ and then take its time derivative. Using the q-Schr\"odinger in eq. (\ref{qschrodinger2}) and its complex conjugate equation we arrive at 
\begin{equation}
\label{towardscontinuity1}
i \hbar \partial_t \rho = \frac{\hbar^2 }{2m} \left[  \Psi (\nabla^2 \Psi^{*}) - \Psi^{* }  (\nabla^2 \Psi)        \right].
\end{equation}
Now, using $\vec{\nabla}.(\varphi \vec{\nabla} \chi) = \vec{\nabla} \varphi . \vec{\nabla} \chi + \varphi \nabla^2 \chi$, firstly for $\varphi = \Psi \; \text{;} \; \chi = \Psi^{*}$ and  in the second case for $\varphi = \Psi^{*}\text{;} \; \chi = \Psi$. Subtracting the second case from the first we get
the generalized equation of continuity
\begin{equation}
\label{continuity}
\partial_t \rho + \vec{\nabla} . \vec{j} = 0 , 
\end{equation}
where
\begin{equation}
\label{corrente}
\vec{j} = \frac{\hbar}{2im} \left[      \Psi^{*} (\vec{\nabla} \Psi) -  \Psi (\vec{\nabla} \Psi^{*} )          \right]
\end{equation}
and $\rho = (\Psi^{*} \Psi)^q$.
For the model in \cite{nrt} to have an equation of continuity it is necessary a constraint to be verified by its solution \cite{nrt2}, instead the present model has an equation of continuity for every field $\Psi$ which is  a solution of the model.
Although the form of eqs. (\ref{continuity}) and (\ref{corrente}) are the same as in the undeformed case, note that its meaning is different since the above definition of $\rho$ carries the parameter $q$ and the wave function $\psi$ is a solution of eq. (\ref{qschrodinger2}) which is different from the standard case.

\subsection{Perturbative Solution}
\label{perturbative}
We are going to solve the $q$-Schr\"odinger equation given in eq. (\ref{qschrodinger2}) for $V(\vec{x})=0$ to first order in $\epsilon\equiv1-q$. We define as usual $z\equiv i(\vec{k}.\vec{x}-\omega t)$. In terms of $z$ and $\epsilon$ the $q$-Schr\"odinger becomes
\begin{equation}
\label{qschrodinger3}
\hbar \omega (1-\epsilon) \Psi' - \frac{\hbar^2 \vec{k}^2}{2 m}  (\Psi^{*}\Psi)^{\epsilon} \Psi'' = 0 ,
\end{equation}
where the derivatives are taken with respect to the variable $z$. The constants are eliminated by using $\omega = \frac{\hbar \vec{k}^2}{2 m}$.

We expand $\Psi$ order to order in $\epsilon$ as $\Psi = \Psi_0 + \epsilon \Psi_1 + ... +\epsilon^n \Psi_n + ...$. To first order the expansion of the two terms of eq. (\ref{qschrodinger3}) is
\begin{eqnarray}
\label{t1}
(1-\epsilon) \Psi'  &=&  \Psi_0' + \epsilon \left(   \Psi_1' - \Psi_0'     \right)  + O(\epsilon^2)   ,  \\
\label{t3}
 (\Psi^{*} \Psi)^{\epsilon} \Psi''  &=&    \Psi_0'' + \epsilon \left[  \Psi_1'' + \Psi_0''  \log   \left( \Psi_0  \Psi^{*}_0  \right)    \right]     + O(\epsilon^2)   .              
\end{eqnarray}
To zeroth order in $\epsilon$ we find the standard equation, $\Psi_0' = \Psi_0''$, with solution $\Psi_0(z) = \exp(z)$. To first order in $\epsilon$ the equation is 
$\Psi_1' - \Psi_1'' = \exp(z)$, with solution
\begin{equation}
\label{solutionepsilon}
\Psi_1(z) = \exp(z) (1-z+c_1) + c_2 .
\end{equation}
Thus, to first order in $\epsilon$ the solution is
\begin{equation}
\label{solutiontoepsilon}
\Psi(z) = \exp(z) \left[  1 + \epsilon \left(     (1-z+c_1) + c_2         \right)          \right]
\end{equation}
 
It is interesting to notice that both equations are linear. The to zeroth order in $\epsilon$ is an homogeneous equation while to first order in $\epsilon$ is a non-homogeneous one. This pattern seems to repeat for orders higher than zero in $\epsilon$.

We can easily show that for a general $V(\vec{x})$ the equation we must solve for the first order contribution is
\begin{equation}
\label{epsilonequation}
i \hbar  \partial_t \Psi_1 + \frac{\hbar^2}{2 m} \nabla^2 \Psi_1 - V(\vec{x}) \Psi_1 = i \hbar \left[ 1+ \log(\Psi^{*}_0 \Psi_0)  \right] \partial_t \Psi_0 ,
\end{equation} 
where $\Psi_0$ is the solution of the standard, non-deformed, Schr\"odinger equation with potential $V(\vec{x})$. Again, to zeroth order in $\epsilon$ is an homogeneous equation while the equation to first order in $\epsilon$ is a non-homogeneous one. It would be interesting to find the first-order solution in epsilon for a physical potential in order to understand the result of deformation in a physical system.

\section{Deformed Hamiltonian}
\label{sechamiltonian}

In order to construct the deformed Hamiltonian from the deformed Lagrangian given in eq. (\ref{qlagrangian}) it is convenient to generalize the correspondence between a field and a generalized coordinate in classical mechanics. For simplicity, we discretize this deformed system. We divide the space $S$ where the system is defined into $N$ small cubes where each cube has volume $v$. The value of $\Psi(\vec{x},t)$ in each tiny cube is defined as $\Psi(\vec{x}_i)\equiv \Psi_i(t)$ where $\vec{x}_i$ is the coordinate of a point arbitrarily chosen in this small cube. 

Let us focus on the time dependent term of the Lagrangian in eq. (\ref{qlagrangian})
\begin{equation}
\label{timeqlagrangian2}
L = \frac{i \hbar}{q} \int d^3 x  \Psi^{* q} \partial_t \Psi^q  + \dots ,
\end{equation}
where $\dots$ corresponds to the $t$-independent terms.
We define the generalized coordinate associated to the discretized field $\Psi_i(t)$ as
\begin{equation}
\label{qcorrespondence}
r_i(t) \equiv v \Psi_i(t)^q  .
\end{equation}
If we did not make this generalization we would find a wrong value of the conjugate momentum.

Thus, the discretization of eq. (\ref{timeqlagrangian2}) is
\begin{equation}
\label{discretization}
L = \frac{i \hbar}{q} \sum_{j} v \Psi^{*}_j (t)^q \frac{\dot{r}_j(t)}{v} + \dots  .
\end{equation}
From the above equation we compute the canonically conjugate momentum associated to $r_j(t)$ as
\begin{equation}
\label{conjmomentum}
p_j(t)  = \frac{\partial L}{\partial \dot{r}_j (t)} = \frac{i \hbar}{q} \Psi^{*}_j (t)^q .
\end{equation}
With the above value of $p_j(t)$ the Hamiltonian can be easily computed in the standard way
\begin{eqnarray}
\label{discretizedhamiltonian}
H = \sum_j p_j \dot{r}_j - L =\frac{i \hbar}{q} \sum_j \Psi_j^{*} (t)^q  v \frac{d{\Psi}_j(t)^q}{dt} -L .
\end{eqnarray}
When $v \rightarrow 0$ we find the Hamiltonian as
\begin{equation}
\label{hamiltonian}
H = \int d^3 x \left[   \frac{\hbar^2}{2 m} \vec{\nabla} \Psi^{*}  .  \vec{\nabla} \Psi + V(\vec{x})   \Psi^{*} \Psi   \right] .
\end{equation} 
It is clear that the above Hamiltonian becomes the standard Hamiltonian for the Schr\"odinger equation as $q \rightarrow 1$. 
The $\Psi$ wave function in equation (\ref{hamiltonian}) is a solution to equation (\ref{qschrodinger2}), so it implicitly depends on the parameter $q$.

Since
\begin{equation}
\lim_{q \rightarrow 1} \mathbb{D}^{(q)}_x f(x) = \frac{d f(x)}{dx},
\end{equation}
we trivially see from all deformed expressions that we obtain the standard expressions as the deformation parameter $q \rightarrow 1$.

\subsection{Conservation of Energy}
\label{conservationen}
In this subsection, following standard steps, we are going to show that the energy is conserved. The density of energy is
\begin{equation}
\label{properdensityen}
\mathcal{H} = \frac{\hbar^2}{2 m} \vec{\nabla} \Psi . \vec{\nabla} \Psi + V(\vec{x})   \Psi^{*} \Psi   , 
\end{equation}
and the the $q$-Schr\"odinger equation is
\begin{equation}
\label{properequation}
i q \hbar \partial_t \Psi + \frac{\hbar^2}{2m} f \nabla^2 \Psi - V f \Psi = 0 ,
\end{equation}
where the factor $f$ in eq. (\ref{properequation}) is
\begin{equation}
\label{ffactor}
f \equiv (\Psi^{*}\Psi)^{1-q} .
\end{equation}
Now, taking the derivative of eq. (\ref{properdensityen}) with respect to $t$ and using eq. (\ref{properequation}) and its complex conjugate equation we have
\begin{eqnarray}
i q \hbar \partial_t \mathcal{H} = \left(\frac{\hbar^2}{2 m}\right)^2 \left[ \vec{\nabla} \left( f \nabla^2 \Psi^{*} \right).\vec{\nabla}\Psi  -    
\vec{\nabla} \Psi^{*} . \vec{\nabla} \left(  f \nabla^2 \Psi     \right) \right]   \nonumber \\ 
+   \frac{\hbar^2}{2 m} \left[ -\vec{\nabla} \left( V f \Psi^{*} \right) . \vec{\nabla} \Psi  + \left( \vec{\nabla} \Psi^{*}  \right) . \vec{\nabla} \left( V f \Psi  \right)            \right] \nonumber \\
\label{intermediate1}
+   \frac{\hbar^2}{2 m} V f \left[  \left( \nabla^2 \Psi^{*} \right) \Psi - \Psi^{*} \nabla^2 \Psi            \right]
\end{eqnarray}

Firstly, we work on the first two terms of eq. (\ref{intermediate1}). Now, we use the identity 
\begin{equation}
\label{identity}
\vec{\nabla}.(\phi \vec{\nabla}\chi)=\phi \nabla^2\chi + \vec{\nabla}\phi.\vec{\nabla}\chi 
\end{equation}
for
\begin{eqnarray}
\label{number1}
\text{(1)} \, \chi = \Psi  \; \;    \text{and}  \; \;   \phi = f\nabla^2\Psi^{*},   \\
\label{number2}
\text{(2)} \, \chi = \Psi^{*}  \; \;   \text{and}   \; \;   \phi = f\nabla^2\Psi ,
\end{eqnarray}
and thus the use of identity in eq. (\ref{identity}) for the case in eq. (\ref{number1})  minus the identity for eq. (\ref{number2}) leads us to
\begin{eqnarray}
\label{intermediate2}
\vec{\nabla} \left( f \nabla^2 \Psi^{*} \right).\vec{\nabla}\Psi  -   \vec{\nabla} \Psi^{*} . \vec{\nabla} \left(  f \nabla^2 \Psi     \right) =  \nonumber \\
\vec{\nabla} . \left[   f (\nabla^2\Psi^{*}) (\vec{\nabla} \Psi) - (\vec{\nabla} \Psi^{*}) f \nabla^2 \Psi             \right] .
\end{eqnarray}
Now, we work on the third, fourth, fifty and sixty terms of eq. (\ref{intermediate1}). We distribute the gradient of the terms $\vec{\nabla} \left( V f \Psi^{*} \right)$ and $\vec{\nabla} \left( V f \Psi  \right)$ which are in the third and fourth terms in eq. (\ref{intermediate1}). Putting together the positive terms after the distribution of the gradient we are able to write them as a divergence term
\begin{eqnarray}
\label{positive}
\Psi \vec{\nabla} \Psi^{*} .\vec{\nabla} (Vf) + Vf \left( \vec{\nabla} \Psi^{*} . \vec{\nabla} \Psi + \Psi \nabla^2 \Psi^{*}       \right)  \nonumber \\ =
\vec{\nabla} \left( V f \Psi \vec{\nabla} \Psi^{*}         \right) .
\end{eqnarray}
The same procedure is done for the negative terms. Then, we have a equation of continuity for the density of energy $\partial_t \mathcal{H} + \vec{\nabla}.\vec{j}=0$ where
\begin{eqnarray}
\label{current2}
\vec{j} = \frac{\hbar^3}{4 i q m^2} f \left[  (\vec{\nabla} \Psi^{*}) (\nabla^2 \Psi) -  (\nabla^2 \Psi^{*})  (\vec{\nabla} \Psi)           \right]  +  \nonumber \\
\frac{\hbar}{2 i q m} V f \left[ \Psi^{*} \left( \vec{\nabla}\Psi \right)  - \left( \vec{\nabla} \Psi^{*} \right) \Psi              \right].
\end{eqnarray}

\subsection{Conservation of Momentum}
\label{conservationmom}
In deriving the conservation of momentum we will make use of the Lagrangian given by eq. (\ref{qlagrangian2}) and from Schr\"odinger in eq. (\ref{qschrodinger2}) for $V(\vec{x}) = 0$.

As usual the $\vec{x}_j$ component of momentum density is obtained from the Lagrangian as
\begin{equation}
\label{defmomdensity}
\mathcal{P}_j = \frac{\delta \mathcal{L}}{\delta \partial_t \Psi} \partial_{x_j} \Psi + \frac{\delta \mathcal{L}}{\delta \partial_t \Psi^{*}} \partial_{x_j} \Psi^{*}.
\end{equation}
Applying this definition for the Lagrangian in eq. (\ref{qlagrangian}) gives us
\begin{equation}
\label{qxjdensity1}
\mathcal{P}_j = \frac{i \hbar}{q} \Psi^{*q} \partial_{x_j} \Psi^q .
\end{equation}
Taking the time derivative of the above $\vec{x}_j$ component of momentum density and using the $q$-Schr\"odinger in eq. (\ref{qschrodinger2}) for $V(\vec{x})=0$ we get
\begin{equation}
\label{tderivativedensitymom1}
\partial_t  \mathcal{P}_j = \frac{\hbar^2}{2 q m} \left[   \left( \nabla^2 \Psi^{*} \right) \Psi^{1-q} \partial_{x_j} \Psi^q  - \Psi^{*q} \partial_{x_j} \left( \Psi^{*(1-q)} \nabla^2 \Psi     \right)        \right] .
\end{equation}
Using in the first term $\Psi^{1-q} \partial_{x_j} \Psi^q = q \partial_{x_j} \Psi$ and taking the derivative in the second term we get
\begin{eqnarray}
\label{tderivativedensitymom2}
\partial_t  \mathcal{P}_j = \frac{\hbar^2}{q 2m} \left[  q\left(\nabla^2 \Psi^{*}\right)  \partial_{x_j} \Psi - \Psi^{*} \partial_{x_j} \left(  \nabla^2 \Psi     \right)                     
- (1-q)  \left( \partial_{x_j} \Psi^{*}  \right) \nabla^2 \Psi             \right] .
\end{eqnarray}
Adding and subtracting the term $q \Psi^{*} \partial_{x_j} \left(  \nabla^2 \Psi \right)$ into the brackets of the above expression we have
\begin{eqnarray}
\label{tderivativedensitymom3}
\partial_t  \mathcal{P}_j = \frac{\hbar^2}{q 2 m} \left[ q  \left[ \left( \nabla^2 \Psi^{*} \right) \left( \partial_{x_j} \Psi \right) - \Psi^{*} \partial_{x_j} \left( \nabla^2  \Psi \right)                     \right]                 
 - (1-q)    \partial_{x_j} \left( \Psi^{*}  \nabla^2 \Psi    \right)                                                         \right] .  
\end{eqnarray}          
Using the identity $\rho \nabla^2 \varphi - \varphi \nabla^2 \rho = \vec{\nabla} . \left( \rho \vec{\nabla} \varphi -\varphi \vec{\nabla} \rho \right)$ for $\rho = \partial_{x_j} \Psi $ and $\varphi = \Psi^{*}$  we arrive at
\begin{equation}
\label{continuitymomentum}
\partial_t \mathcal{P}_j + \vec{\nabla} . \vec{j} + \frac{(1-q)}{q} \partial_{x_j} K(\vec{x},t) = 0 ,
\end{equation}
where
\begin{equation} 
\label{currentmomentum}
\vec{j} =    \Psi^{*} \left( \vec{\nabla} \partial_{x_j} \Psi \right) -  \left(  \partial_{x_j} \Psi \right) \left(  \vec{\nabla} \Psi^{*}   \right)      
\end{equation}  
and   $K(\vec{x},t) =- \frac{\hbar^2}{2 m} \Psi^{*} \nabla^2 \Psi$ is the density of kinetic energy.    

We thus see from eq. (\ref{continuitymomentum}) that the momentum is conserved with an additional term proportional to $1-q$ which mesures for each component the difference of  kinetic energy at plus and minus infinite which is zero in these limits.

\section{Numerical Computation of the Probability Density}
\label{numerical}
The purpose of this section is to write the $q$-Schr\"odinger equation we have developed here which is given in eq. (\ref{qschrodinger2}) in terms of $\rho$, which is a generalized probability density. In terms of  $\xi = \vec{k}.\vec{x}- \omega t$ the equation is
\begin{equation}
\label{qschrodinger4}
-i q  \Psi' +  \left( \Psi^{*} \Psi \right)^{1-q} \Psi'' - \hat{V} (\Psi^{*} \Psi)^{1-q} \Psi = 0 ,
\end{equation}
where $\hat{V}=V/(\hbar \omega)$, $\Psi' = d\Psi/d\xi$, $\Psi'' = d^2 \Psi/d\xi^2$ and we have considered that $\omega = E/\hbar = (\hbar \vec{k}^2)/(2 m))$.

We write the wave function as $\Psi(\xi) = R(\xi) \exp(i \theta(\xi))$. But according to the equation of continuity, $\rho = (\Psi^{*} \Psi)^{q}$, then the wave function can be written as 
\begin{equation}
\label{wavefunctionprobability}
\Psi = \rho^{\frac{1}{2 q}} \exp(i \theta).
\end{equation}
It is straightforward to express the three terms of eq. (\ref{qschrodinger4}) in terms of $\rho$ and $\theta$ and the result is
\begin{eqnarray}
\label{qthreeterms}
-\frac{i}{2} \rho^{\frac{1-2q}{2q}} \rho'    + q    \rho^{\frac{1}{2q}} \theta'    +    \frac{1-2q}{4q^2} \rho^{\frac{3-6q}{2q}}  \left( \rho' \right)^2    -    \rho^{\frac{3-2q}{2q}} \left( \theta' \right)^2  +                \nonumber                 \\
+ \frac{i}{q} \rho^{\frac{3-4q}{2q}} \rho' \theta' + \frac{1}{2q} \rho^{\frac{3-4q}{2q}} \rho'' +  i \rho^{\frac{3-2q}{2q}} \theta'' - \hat{V} \rho^{\frac{3-2q}{2q}} = 0 .                             
\end{eqnarray}   
The above equation shows imaginary and real terms. Putting apart the imaginary and real terms we have two coupled equations as shown below
\begin{eqnarray}
\label{imaginary}
-\frac{1}{2} \rho^{\frac{1-2q}{2q}} \rho' +  \frac{1}{q} \rho^{\frac{3-4q}{2q}} \rho' \theta'  +   \rho^{\frac{3-2q}{2q}} \theta''  = 0  ,   \\
\label{real}
q \rho^{\frac{1}{2q}} \theta'  + \frac{1-2q}{4q^2} \rho^{\frac{3-6q}{2q}}  \left( \rho' \right)^2  - \rho^{\frac{3-2q}{2q}} \left( \theta' \right)^2 +  \frac{1}{2q} \rho^{\frac{3-4q}{2q}} \rho'' - \hat{V} 
\rho^{\frac{3-2q}{2q}} = 0 
\end{eqnarray} 
which must be solved. 

Using an algebraic computer software we solved numerically the set of equations in eqs. (\ref{imaginary} and \ref{real}) for $V(\vec{x})=0$. We present the result of the numerical computation in figures for eight values of $q$ of the function $\rho$ which shows the behavior of the function $\rho$ as the decreasing parameter $q$ goes from $q \geq 1$ to  $q \leq 0$.

\begin{figure}
\begin{center}
\includegraphics[width=2in]{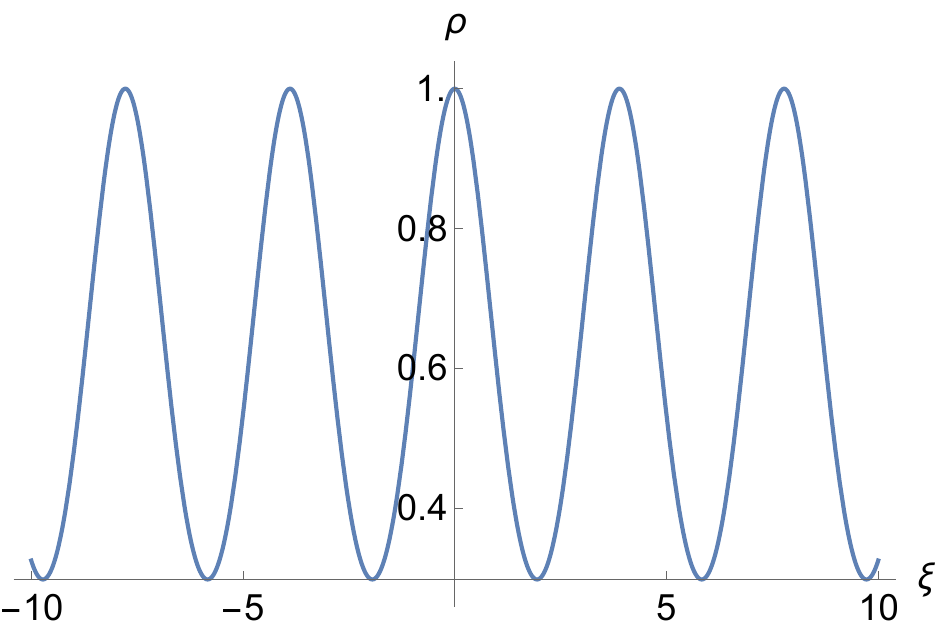}
\caption{ $\rho$ versus $\xi = \vec{k}.\vec{x}-\omega t$  for $q = 1.4$.}
\label{densityq1p4}
\end{center}
\end{figure}

\begin{figure}
\begin{center}
\includegraphics[width=2in]{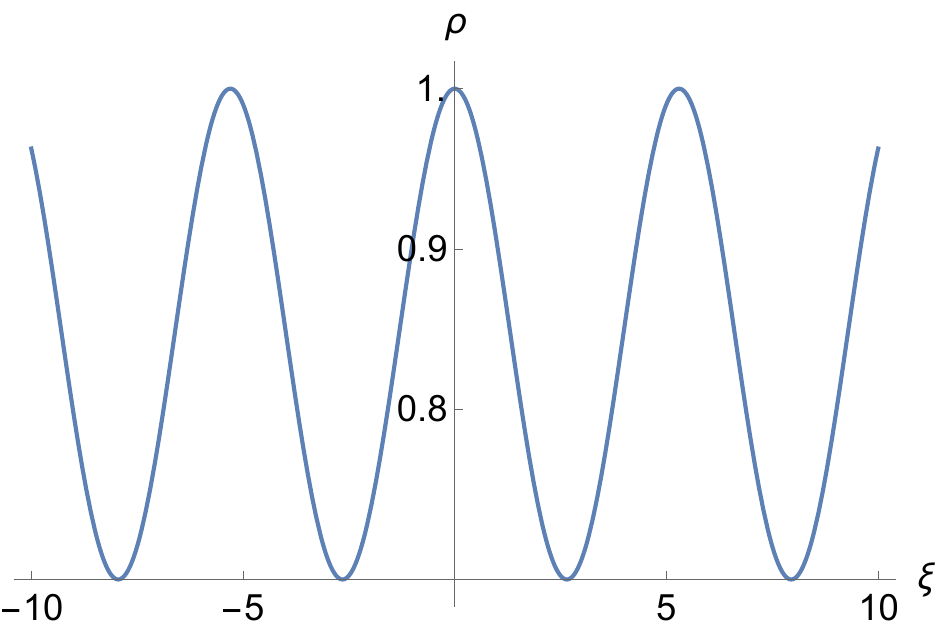}
\caption{ $\rho$ versus $\xi = \vec{k}.\vec{x}-\omega t$  for $q = 1.1$.}
\label{density}
\end{center}
\end{figure}

\begin{figure}
\begin{center}
\includegraphics[width=2in]{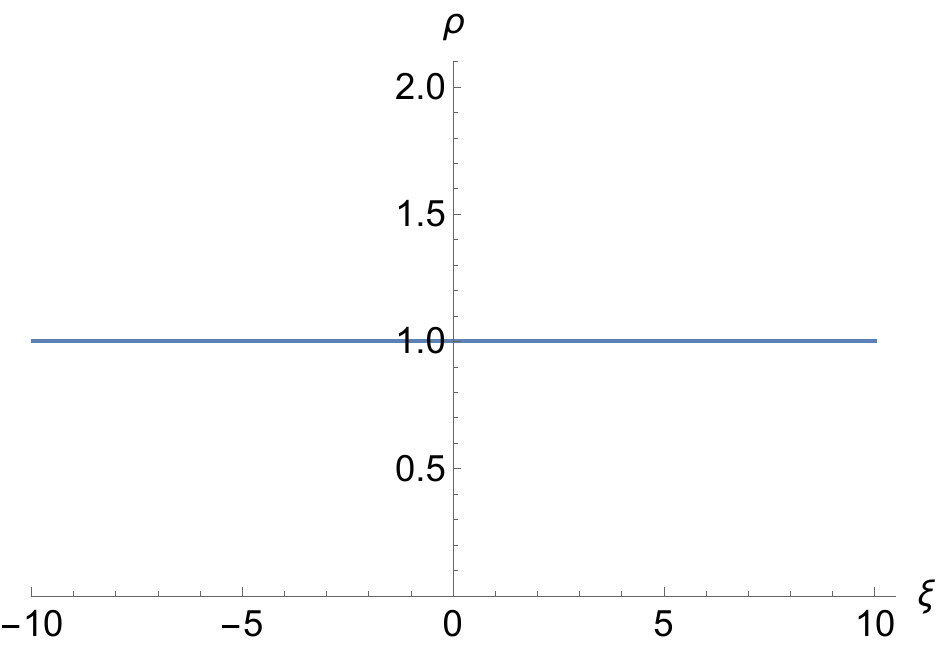}
\caption{ $\rho$ versus $\xi = \vec{k}.\vec{x}-\omega t$  for $q = 1$.}
\label{density2}
\end{center}
\end{figure}

%
%


\begin{figure}
\begin{center}
\includegraphics[width=2in]{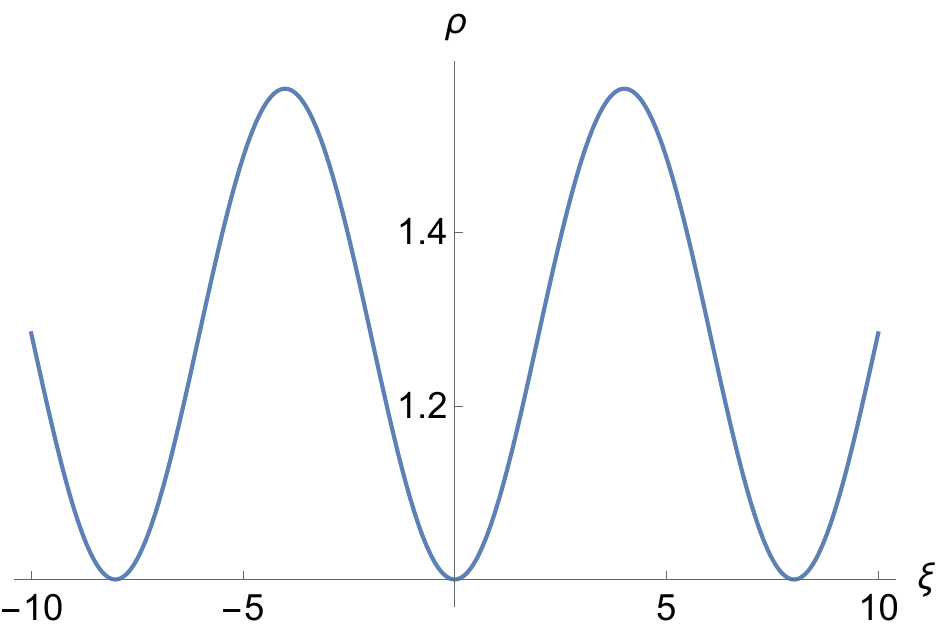}
\caption{ $\rho$ versus $\xi = \vec{k}.\vec{x}-\omega t$ for $q = 0.9$.}
\label{density3}
\end{center}
\end{figure}

\begin{figure}
\begin{center}
\includegraphics[width=2in]{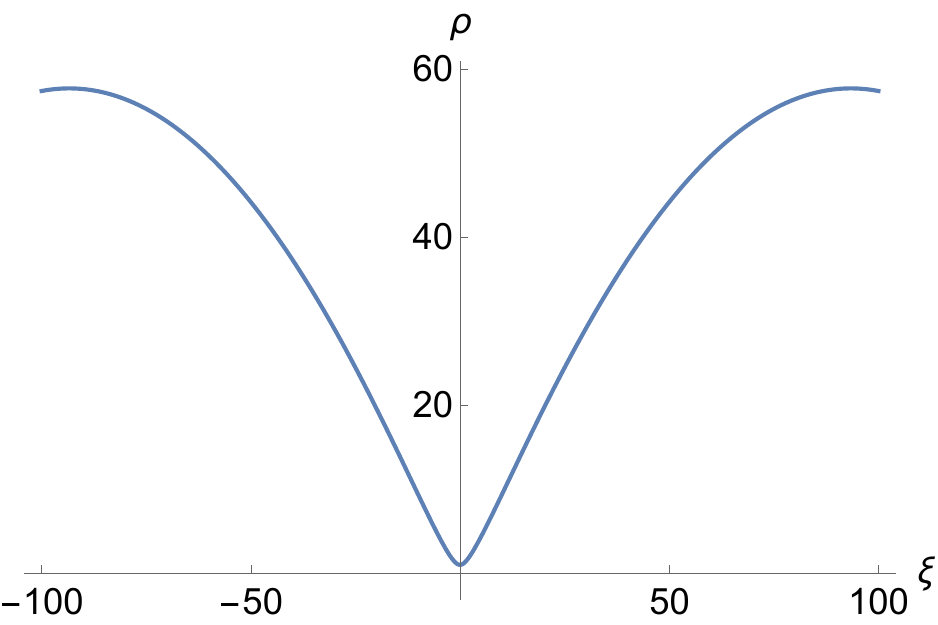}
\caption{ $\rho$ versus $\xi = \vec{k}.\vec{x}-\omega t$  for $q = 0.6$.}
\label{density4}
\end{center}
\end{figure}

\begin{figure}
\begin{center}
\includegraphics[width=2in]{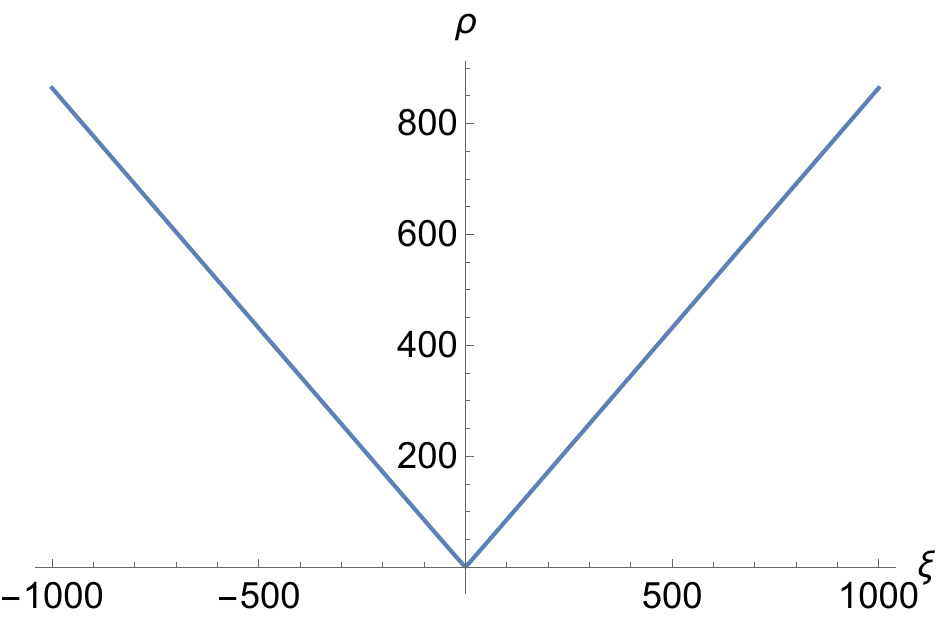}
\caption{ $\rho$ versus $\xi = \vec{k}.\vec{x}-\omega t$  for $q = 0.5$.}
\label{density5}
\end{center}
\end{figure}

\begin{figure}
\begin{center}
\includegraphics[width=2in]{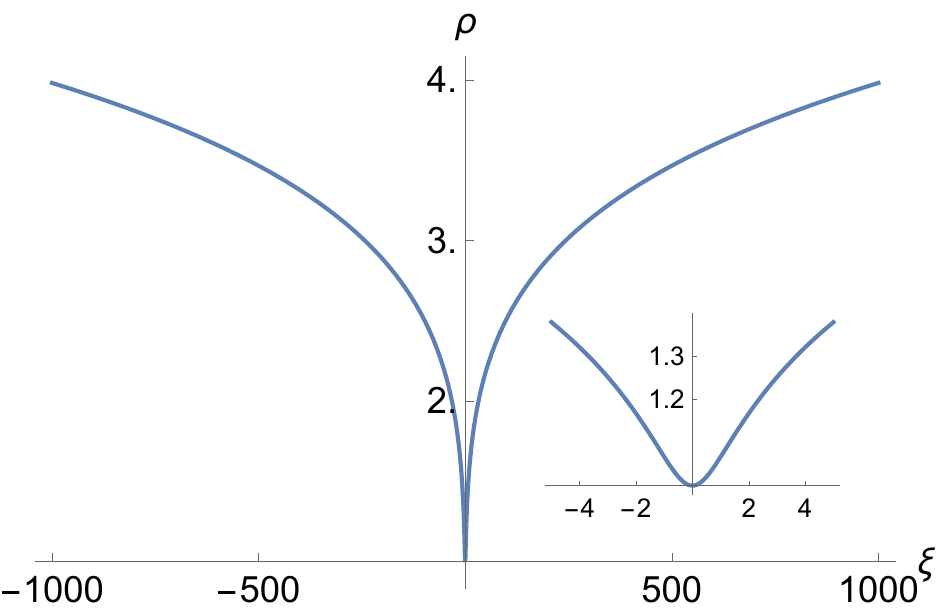}
\caption{ $\rho$ versus $\xi = \vec{k}.\vec{x}-\omega t$  for $q = 0.1$. In the inset it is possible to see, in a smaller scale, that there is no cusp at zero.}
\label{density6}
\end{center}
\end{figure}

 

\begin{figure}
\begin{center}
\includegraphics[width=2in]{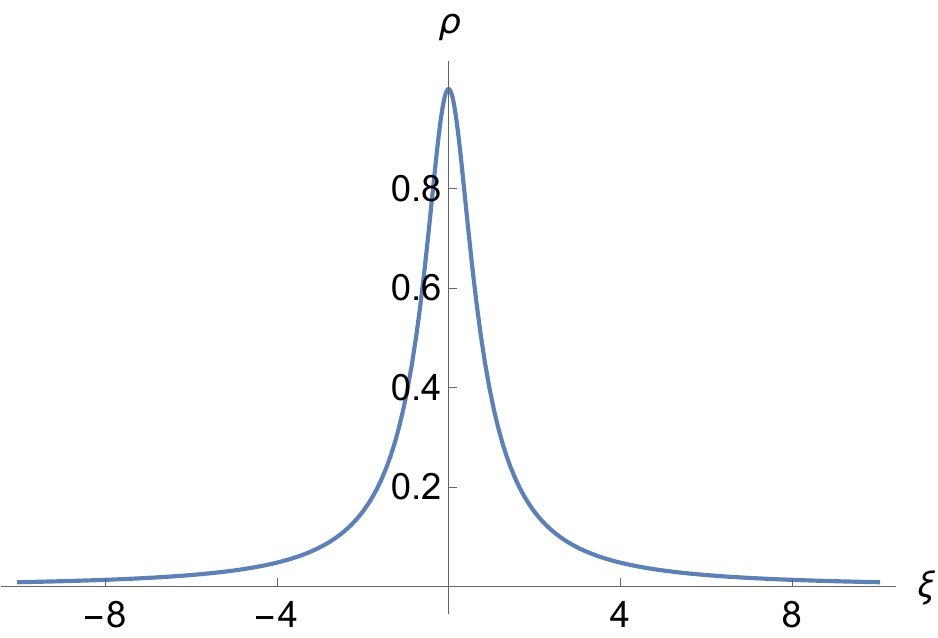}
\caption{ $\rho$ versus $\xi = \vec{k}.\vec{x}-\omega t$  for $q = -1.0$.}
\label{density7}
\end{center}
\end{figure}



\vspace{1cm}

We would like to make an interesting remark here. Notice that, in fig. \ref{densityq1p4}, for $q = 1.4$, the 
wavelength of the oscillations ($\rho \in [0,1]$) is smaller than the wavelength for $q=1.1$, fig. \ref{density}. This 
implies that the wavelength increases when $q$ decreases, approaching $q=1$ from 
above.

In fact, when $q=1$ 
the wavelenth of the oscillations diverges, since the solution of eq. (\ref{qschrodinger2}) 
(or eq. (\ref{qschrodinger4}))  for $q=1$ is the plane wave with  $\rho = 1$,
fig. \ref{density2}. 

This scenario looks much like the behavior 
of the correlation length of a ferromagnetic material  when we increase the temperature 
from lower temperature (inside the ordered phase) to the critical temperature ($T_c$) in the 
ferromagnetic-paramagnetic  phase transition. The correlation length of the finite clusters 
increases when we increase  the temperature, from 
lower temperatures, and diverges at 
$T = T_c$, exactly as the wavelength of $\rho$ does. Also, in the ferromagnetic-paramagnetic 
continuous phase transition, there is a change in symmetry, from an ordered phase where 
the symmetry of rotation does not exists, due the existence of a magnetization, to a disordered phase 
where this symmetry does exists, since the magnetization is zero, i.e., there is no privileged direction. 
Here, we have a similar scenario. The parameter $q$ plays the role of inverse temperature. 
The region $q > 1$ corresponds to the ordered ferromagnetic phase, and decreasing $q$ 
corresponds to increasing  the temperature in a ferromagnetic system that is in the ordered phase.
We also have here, at $q=1$, restoring a symmetry that does not exist in the region $q>1$. 
At exactly $q=1$, we have the symmetry of arbitrary translations, symmetry that does not exists in 
the region corresponding to $q>1$, where there is the symmetry of discrete translations, corresponding 
to the wavelength. 
So, the correlation length and the wavelength of $\rho$ present this similar scenario. It is clear 
that the integration of $\rho$ over the whole space diverges, as the plane wave. 

The scenario for $q<1$ shows that, for $0.5 < q < 1$, the oscillations reappear, but the 
oscillations are always above one in the ordinate ($\rho \geq 1$),  
see figs. \ref{density3} and \ref{density4}. 
The wavelengths of these oscillations, as well as their amplitudes,  
increase when $q$ goes from $q$ slightly smaller than one  to $q$ slightly greater than $0.5$ and 
at $q = 0.5$, fig. \ref{density5},  the amplitude goes to infinity and there is no more periodicity. 
If we continue decreasing the value of $q$, below $q=0.5$, the function $\rho$ increases  monotonically, 
but it grows less rapidly as the $q$ parameter decreases, 
without oscillations, see figs. \ref{density5} and \ref{density6}. Of course, the integration on space of all these densities having $q > 0$ diverges.

At exactly $q=0$ we recover again the plane wave, having $\rho = 1$. 

The scenario for $q < 0$ is completely different. Interesting soliton solutions appear, without any oscillations, fig. \ref{density7} for $q = -1.0$. The integration of $\rho$ over the whole space is, in this case, finite. 
This situation is odd with all scenarios with $q \ge 0$. Clearly, this solitonic solutions have appealing  
physical consequences that deserve future investigations. 

\vspace{1cm}

\section{Final Comments}
\label{final}

We construct a nonlinear deformed model of the Schr\"odinger structure by modifying the kinetic term by introducing a parameter $q$ used in the power of the wave function of the kinetic term. This model has a standard potential energy $V(\vec{x})$. It has also a classical field Lagrangian and Hamiltonian which can interact with light and it is shown that the generalization of the probability density obeys an equation of continuity. Moreover, energy and momentum are shown to be conserved. For $q=1$ the model reproduces trivially the standard Schr\"odinger structure.

For the case of a free particle, $V(\vec{x})=0$, we have solved the qNLSE by a perturbation expansion to first order in $\epsilon=1-q$. It is interesting to mention that the equations that appear in first and second order in $\epsilon$ are linear. The zeroth order in $\epsilon$ is an homogeneous equation while to first order in $\epsilon$ is a non-homogeneous one. This pattern seems to repeat for orders higher in $\epsilon$. 

We have also computed numerically the $\rho$ for the case of $V(\vec{x})=0$ in any spatial dimensions and for arbitrary values of $q$. The result shows a solitonic behavior when the space is one and $q < 0$. When $q= 0,1$ $\rho$ is constant elsewhere accordingly to the standard case and in the intervals 
$q > 0.5$ (excluding $q=1$) $\rho$ is wavy,  and for $0 < q < 0.5$ $\rho$ is a monotonically increasing 
function, without periodicity.

It is interesting to compare this model with the previous model presented in \cite{nrt} and \cite{nrt2}. The disadvantage of the present model compared to the previous one is that we have only been able so far to solve numerically the deformed equation in the case of the free particle. The advantage of this model over the previous one is threefold. The first is that only one field is necessary for this model to describe the system, while for the previous model it is necessary to introduce an additional field to describe the system correctly. The second point is that this model can interact with the electromagnetic field. Moreover, differently from the previous model the present model has an equation of continuity for every solution of the model as is shown in eq. (\ref{continuity}). It is shown in \cite{nrt2} that only the solutions satisfying a constraint lead to conservation of probability. Finally, the probability density of the previous model is constant everywhere in any space dimension, as in the case of the standard Schr\"odinger equation, while in the present model the probability density is wavy in the intervals $0 < q < 1$ and $q > 1$ and presents a solitonic behavior for $q < 0$ for space equals to one.

\vspace{1cm}

\noindent

\vspace{2cm} 
\noindent 
{\bf Acknowledgments:}   The authors thank Constantino Tsallis for discussions. One of us, EMFC,  acknowledges CNPq and FAPERJ for financial support.

\end{document}